\documentclass{article}
\usepackage[numbers]{natbib}
\usepackage{hyperref}
\usepackage{amsmath}
\usepackage{amssymb}
\usepackage{dsfont}
\usepackage{amsthm}

\usepackage{algpseudocode}
\usepackage{algorithm}

\usepackage{caption}
\usepackage{subcaption}
\usepackage{graphicx} 

\algnewcommand\algorithmicinput{\textbf{Input:}}
\algnewcommand\algorithmicparameter{\textbf{Parameters:}}
\algnewcommand\algorithmicoutput{\textbf{Output:}}
\algnewcommand\Input{\item[\algorithmicinput]}%
\algnewcommand\Parameter{\item[\algorithmicparameter]}%
\algnewcommand\Output{\item[\algorithmicoutput]}%

\hypersetup{colorlinks=true, citecolor = blue, linkcolor = red}

\title{Innovative Non-parametric Texture Synthesis via Patch Permutations}

\author{\parbox{\textwidth}{\centering Ryan Webster\thanks{Code for algorithms and figures is provided on \href{https://github.com/ryanwebster90/OT-texture-synthesis}{github}. All images were from the arbyreed texture dataset found
\href{https://drive.google.com/drive/folders/0B6oh_CUacdkDSkR3cDYyZnBaRDA?usp=sharing}{here}. Credit for all images goes to flickr user \href{https://www.flickr.com/photos/19779889@N00/}{arbyreed}. } } }

\begin{document}

\maketitle
\begin{abstract}
   In this work, we present a non-parametric texture synthesis algorithm capable of producing plausible images without copying large tiles of the exemplar. We focus on a simple synthesis algorithm, where we explore two patch match heuristics; the well known Bidirectional Similarity (BS) measure and a heuristic that finds near permutations using the solution of an entropy regularized optimal transport (OT) problem. Innovative synthesis is achieved with a small patch size, where global plausibility relies on the qualities of the match. For OT, less entropic regularization also meant near permutations and more plausible images. We examine the tile maps of the synthesized images, showing that they are indeed novel superpositions of the input and contain few or no verbatim copies. Synthesis results are compared to a statistical method, namely a random convolutional network. We conclude by remarking simple algorithms using only the input image can synthesize textures decently well and call for more modest approaches in future algorithm design.

\end{abstract}  

\section{Introduction}

Non-parametric texture synthesis algorithms, initiated by \cite{efros1999texture}, operate by copying patches from the exemplar to synthesis.  Since then, these methods have improved considerably in terms of their realism, synthesizability domain and spacetime efficiency. Nearly all modern formulations follow the optimization based method of \cite{kwatra2005texture}, where patches are iteratively matched and re-averaged in the synthesis. Such a formulation allows local regions to propagate to global plausibility. Since then, methods have sought to improve synthesizability domain by enforcing various constraints on the match. Notably,  the Bidirectional Similarity method \cite{simakov2008summarizing} promotes matches using all patches from the input. For example, it will not converge on implausible low energy configurations, such as a constant image using a single patch, whereas the nearest neighbor (NN) match in \cite{kwatra2005texture} does in practice. Very recent methods, such as \cite{ortega2017optimal} find optimal permutations of patches using the hungarian algorithm. On a similar note, \cite{ferradans2013regularized} formulates the problem of color transfer as a regularized discrete optimal transport problem, whose solution is computed using linear programming. Also in a similar vein, \cite{tartavel2015variational} forces patch usage statistics in a learned dictionary. In this work, we take a slightly different approach to \cite{ortega2017optimal}, instead obtaining an approximate solution using Sinkhorn's algorithm. This turns out to be critical for synthesizing high resolution images, where the cost matrix does not fit into memory and all computation has to be sliced.

\subsection{Contributions}
We present two contributions in this work. Our first contribution is the demonstration that entropic optimal transport can be used for texture synthesis in Section 2. We do not utilize the transport plan directly, instead using it to estimate a near permutation between input and synthesis patches. Low values of the entropic regularizer yield near permutations in practice. Our second contribution is a demonstration that non-parametric algorithms can produce novel images by using small patch sizes in Section 3. Synthesis results are present in Figure \ref{fig:icts}, alongside the BS matching heuristic and synthesis with a random convolutional network \cite{ustyuzhaninov2016texture}. 

\section{Match Heuristics}
The method of \cite{kwatra2005texture} aims to minimize the following objective
\begin{equation}\label{1}
\min_{y}\min_{\tilde{\sigma}} \ \  \Vert \tilde{\sigma}\mathcal{P}(x) - \mathcal{P}(y) \Vert^{\beta}_{2}
\end{equation}
Here $x$ is the exemplar image as a $N \times N \times 3$ tensor, $y$ is the current synthesis image, $\mathcal{P}$ is the patchifying linear operator and $\tilde{\sigma} \in \{ 0,1 \}^{N^{2}}$ is a $N \times N$ binary matrix such that $\tilde{\sigma}\mathds{1} = \mathds{1}$, i.e. every patch in $y$ has a match in $x$. $\mathcal{P}$ reshapes all $b \times b \times 3$ patches in periodic extensions of $x$ and $y$ into $N^{2} \times 3 b^{2}$ tensors $X$ and $Y$, whose rows contain patches. In \cite{kwatra2005texture} they use $\beta = .8$ to promote image sharpness however in this work we use $\beta = 2$ as it allows a fast parallel computation of distances using matrix multiplication. \cite{kwatra2005texture} solves \eqref{1} by alternating between a solution for $y$, which can be obtained by solving a sparse system and a solution for $\tilde{\sigma}$ obtained with a nearest neighbor search. In practice, this optimization method can fail because there are no constraints upon $\tilde{\sigma}^{\intercal}\mathds{1}$, where for some images the synthesis will reuse a very small set of patches in $\mathcal{P}(x)$ across the synthesis. \cite{simakov2008summarizing} mitigates this, by minimizing an objective similar to the following
\begin{equation}\label{2}
\min_{y}\min_{\tilde{\sigma}} \ \  \alpha \Vert \tilde{\sigma}\mathcal{P}(x) - \mathcal{P}(y) \Vert^{2}_{2} + (1-\alpha)\Vert \mathcal{P}(x) - \tilde{\sigma}^{\intercal}\mathcal{P}(y) \Vert^{2}_{2}
\end{equation}
Here, $\alpha$ is a parameter that balances $y$ and $x$'s nearest neighbor match choice. Both \eqref{1} and \eqref{2} are minimized in a multi resolution framework, as are the majority of texture synthesis algorithms, non-parametric or otherwise. Algorithm \ref{alg:MRTS} represents a vanilla multi-resolution texture synthesis algorithm. 
\begin{algorithm}
\caption{Multi-Resolution Texture Synthesis}
\label{alg:MRTS}
\begin{algorithmic}[1]
\Input Exemplar image $x$, Number of scales $J$, Number of synthesis iterations, Match heuristic
\Output Synthesized texture $y$

\State $y \gets$ Noise sampled from uniform distribution
\For{$j = J \dots 0$}
\State $x_{j} \gets x$ at resolution $2^{-j}$
\State $y_{j} \gets y$ at resolution $2^{-j}$
\For{Number of synthesis iterations}
\State $X \gets \mathcal{P}(x_{j})$ 
\State $Y \gets \mathcal{P}(y_{j})$ 
\State $\tilde{\sigma} \gets $ Result of match heuristic on $X$ and $Y$
\label{algline:mh}
\State $Y \gets \tilde{\sigma}X$
\State $y_{j} \gets $ Re-average $Y$ back into the image domain
\EndFor
\State $y \gets y_{j}$
\EndFor
\end{algorithmic}
\end{algorithm}

For BS \eqref{2}, the match heuristic is performed slightly differently than line \ref{algline:mh}, where instead two separate NN searches are performed and $y$ is updated as a convex combination of the two corresponding updates, according to $\alpha$. In practice, the patchifying operator $\mathcal{P}$ takes a random subset of patches which in this work was $35\%$ for every experiment. Additionally, when using small patch sizes in tandem with sub sampling, high resolutions can drift from the previous low resolution synthesis. To fix this, we re - average the synthesis with lower resolutions at every iteration.  

\par
As we suggested before, \eqref{2} alleviates the short comings of \eqref{1} because it promotes a uniform usage of patches from the exemplar. We define this notion here empirically as the \textit{Match Cardinality}
\begin{equation}\label{3}
MC(\tilde{\sigma}) = \frac{\Vert \tilde{\sigma}^{\intercal}\mathds{1} \Vert_{0}}{N}
\end{equation}
where $ \Vert \tilde{\sigma}^{\intercal}\mathds{1} \Vert_{0}$ is the number of columns in $\tilde{\sigma}$ with at least one nonzero element. In the next section, we'll discuss methods that can achieve permutations or high match cardinalities, using an \textit{optimal transport} approach. 

\subsection{Optimal Transport Formulation}
In recent years, optimal transport (OT) has seen a rich set of applications in computer graphics and machine learning. Recent works \cite{ortega2017optimal} have applied OT to non-parametric texture synthesis. The original optimal transport problem seeks to minimize
\begin{equation} \label{4}
\min_{\Sigma \in \mathcal{S}} \ \langle \Sigma , C_{X,Y} \rangle
\end{equation}
where $\mathcal{S}$ is the set of doubly stochastic matrices  
\begin{equation*}
\mathcal{S} = \{ \Sigma \in {[ 0,1 ]}^{N^{2}} \text{   }| \text{   } \Sigma \mathds{1} = \mathds{1} \text{ and }  {\Sigma}^{\intercal} \mathds{1} = \mathds{1} \}
\end{equation*}
and $C_{X,Y}$ is a distance matrix. In relationship to \eqref{1}, $C_{X,Y}$ represents the euclidean pairwise distances between patches, i.e. ${C_{X,Y}}_{ij} = \Vert X_{i} - Y_{j} \Vert_{2}^{2}$. \eqref{4} is solved exactly with the Hungarian algorithm. \cite{ortega2017optimal} employs this approach to find a global minimum of \eqref{1}, with a $\Vert \cdot \Vert_{2}^{1}$ norm instead of a squared euclidean norm and a permutation $\sigma$ instead of the unconstrained binary matrix $\tilde{\sigma}$. Unfortunately, the Hungarian algorithm runs in $O(N^{3})$ time, which severely limits the resolution of images before the problem becomes intractable. A myriad of approximations which solve \eqref{4} with a permutation (or assignment), for example the auction algorithm \cite{burkard1999linear}. However, for the image resolutions we present in section 5, $C_{X,Y}$ does not fit into memory and has to be computed with a sliced matrix multiplication. In addition, for problems of this size, we'd hope our matching algorithm is parallelized and can run on the gpu. In \cite{deveci2013gpu}, they discuss parallelization schemes for maximum cardinality matching bipartite graphs. They note that in the case where $C$ doesn't fit into memory, these parallelization schemes may be inefficient. For this reason, we turn to a regularized version of \eqref{4}, whose solution can be computed on the gpu and with linear memory. 

\subsection{Entropy Regularized Transport}
Much of the recent attention to OT has been driven by useful approximations to \eqref{4}, most notably the \textit{entropic regularization} schemes initiated by \cite{cuturi2013sinkhorn}. \cite{cuturi2013sinkhorn} proposed modifying \eqref{4} to penalize $\Sigma$ that have low entropy
\begin{equation} \label{5}
\min_{\Sigma \in \mathcal{S}} \ \langle \Sigma , C_{X,Y} \rangle - \varepsilon h(\Sigma)
\end{equation}
where $h(P) = \sum\limits_{ij}-P_{ij}log(P_{ij})$ is the entropy of of matrix $P$. Not only does this convexify \eqref{4}, an optimal solution can be computed extremely efficiently with the Sinkhorn-Knopp (SK) matrix scaling algorithm. What is especially attractive about the algorithm in our scenario, is it can be adapted trivially to a low memory setting, where the matrix $C_{X,Y}$ will not fit into memory and only requires $O(N)$ memory for the scaling vectors. We present this simple modification in Algorithm \ref{alg:LMSK}.

\begin{algorithm}
        \caption{Low Memory Sinkhorn}
        \label{alg:LMSK}
        \begin{algorithmic}
        \Input Row stacked patch matrices $X$ and $Y$
        \Parameter Memory parameter $M$, entropic regularizer $\varepsilon$
        \Output Scaling vectors $a$, $b$
        
        \State $b \gets \mathds{1}_{N}$
        \State $\hat{X}, \hat{Y} \gets$ Concatenate ones and norms of $X,Y$ 
        
        \For{Number of scaling iterations}
        
          \For{Number of slices according to $M$}
              \State $I \gets$ row indices for current slice
              \State $C_{I} \gets \hat{Y}{\hat{X}}^{\intercal}_{I}$
              \State $K_{I} \gets exp(-\frac{C_{I}}{\varepsilon})$
              \State $a_{I} \gets \mathds{1} \oslash ( b_{I}^{\intercal}K_{I})$
          \EndFor

          \For{Number of slices according to $M$}
              \State $I \gets$ row indices for current slice
              \State $C_{I} \gets \hat{Y}_{I}{\hat{X}}^{\intercal}$
              \State $K_{I} \gets exp(-\frac{C_{I}}{\varepsilon})$
              \State $b_{I} \gets \mathds{1} \oslash ( K_{I}a_{I})$
          \EndFor

        \EndFor
        
        \end{algorithmic}
    \end{algorithm}
    
    As is shown in \cite{cuturi2013sinkhorn}, the optimal solution of \eqref{5} is necessarily of the form $\Sigma^{*} = K \otimes (ab^{\intercal})$, where $a,b$ are the output of sufficiently many iterations of the SK algorithm. In our setting, we typically run only a few iterations of Algorithm \ref{alg:LMSK}, implicitly obtaining an approximately doubly stochastic matrix $\Sigma$. The nature of this doubly stochastic matrix $\Sigma$, also called the transport plan, is highly related to the entropic regularizer $\varepsilon$, where high values of $\varepsilon$ represent more uncertainty in the solution, spreading the values of $\Sigma$ across rows and columns. As $\varepsilon \to 0$, $\Sigma$ resembles more of a hard assignment. Assignment promotes a fast convergence of Algorithm \ref{alg:MRTS} and even for small values of $\varepsilon$, using $\Sigma$ directly results in a blurry synthesis as each patch is updated as the convex combination on many input patches. We could define a match in the following way
\begin{equation} 
\delta_{ij} = 
    \begin{cases}
      1, & \text{if}\ \Sigma_{ij} = \max_{1 \leq k \leq N} \{\Sigma_{kj} \} \\
      0, & \text{otherwise}
    \end{cases}
\end{equation}
    
This already works well because it returns a much higher match cardinality than taking minimums on $C$ (i.e. nearest neighbors) and prevents the method of \cite{kwatra2005texture} from outright failing at small patch sizes. A very similar approach was employed in \cite{dufosse2015two} to bipartite graphs, where the adjacency matrix was normalized using SK, and then the columns, which comprise probability distributions, were sampled to provide a match cardinality of $MC = 1 - \frac{1}{e} \approx 0.63$ in expected value. \cite{dufosse2015two} proceeds by using a Karp-Siper heuristic to resolve columns that were not matched uniquely. We take a simpler and more vectorized approach. We resolve columns that were not matched uniquely, i.e. where $\delta^{\intercal}\mathds{1} > 1$, by taking an argument maximum along the rows of $\Sigma$ at the nonzero locations of $\delta$, obtaining a permutation on the support of $\delta$. Iterating this process yields Algorithm \ref{alg:SKP}. Algorithm \ref{alg:SKP} is implemented with the same memory slicing as Algorithm \ref{alg:LMSK} but is presented in this form for brevity. 
    
    \begin{algorithm}
        \caption{Greedy High Cardinality Match}
        \label{alg:SKP}
        \begin{algorithmic}
        \Input $\Sigma \gets K \otimes (ab^{\intercal})$ an approximately doubly stochastic matrix
        \Output A high cardinality match $\tilde{\sigma}$ 
        \State $R,C \gets \{1 \ldots N\}$ sets of unmatched rows and columns
        \For{$k$ = Number of iterations or until desired $MC(\tilde{\sigma})$}
        	\State $\delta \gets$ Arg max of $\Sigma_{R,C}$ along columns as a binary matrix of indices
            \If{$k <$ Number of iterations}
            	\State $\sigma \gets$ Arg max along rows of $\Sigma_{R,C} \otimes \delta$, which is a permutation
                \Else
                \State $\sigma \gets \delta$
            \EndIf
            \State $rs,cs \gets $ Rows and columns indexing the support of $\sigma$
            \State ${\tilde{\sigma}}_{R_{rs},C_{cs}} \gets \sigma_{rs,cs}$ 
            \State $R,C \gets$ Remove matched indices $rs,cs$ from $R$,$C$
        \EndFor
        
        \end{algorithmic}
    \end{algorithm}

\begin{figure}[!ht]
\centering
\hfill
\begin{subfigure}{.24\linewidth}
    \centering
    \includegraphics[width=1\linewidth]{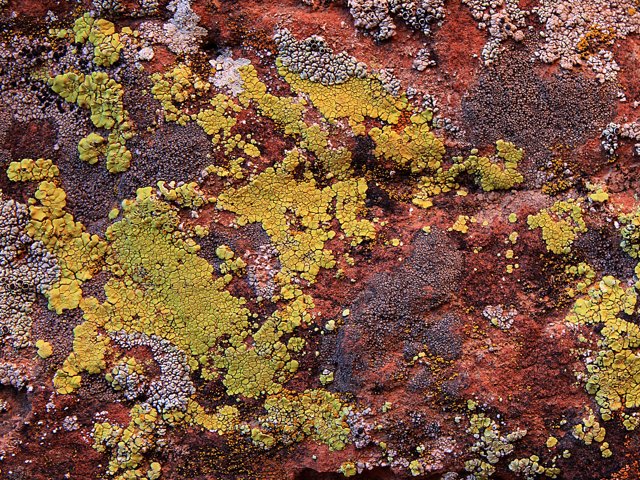}
\end{subfigure}
\begin{subfigure}{.24\linewidth}
    \centering
    \includegraphics[width=1\linewidth]{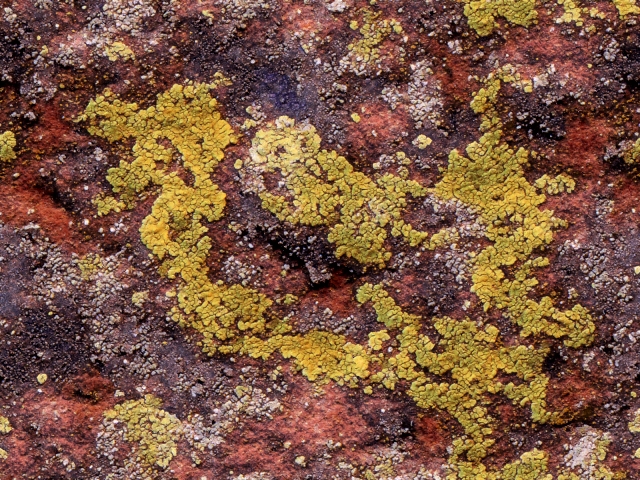}

\end{subfigure}
\begin{subfigure}{.24\linewidth}
    \centering
    \includegraphics[width=1\linewidth]{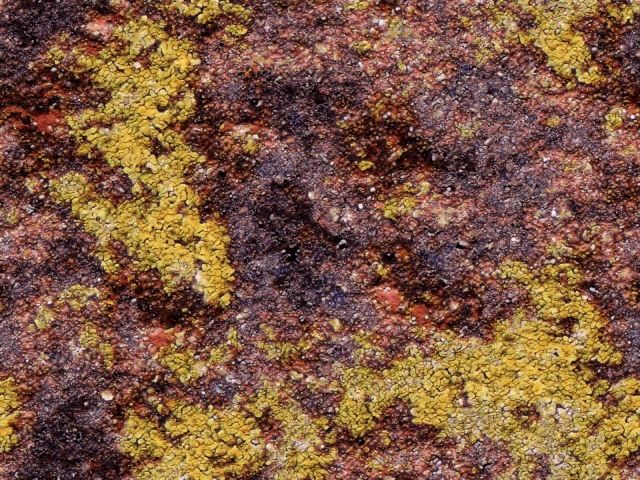}

\end{subfigure}
\begin{subfigure}{.24\linewidth}
    \centering
    \includegraphics[width=1\linewidth]{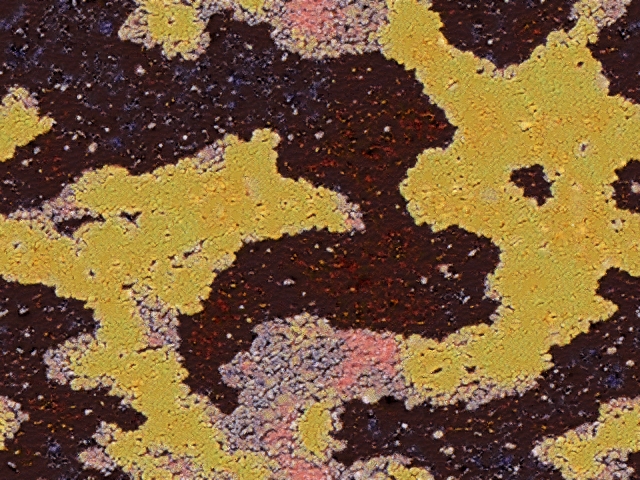}
\end{subfigure}

\smallskip

\hfill
\begin{subfigure}{.24\linewidth}
    \centering
    \includegraphics[width=1\linewidth]{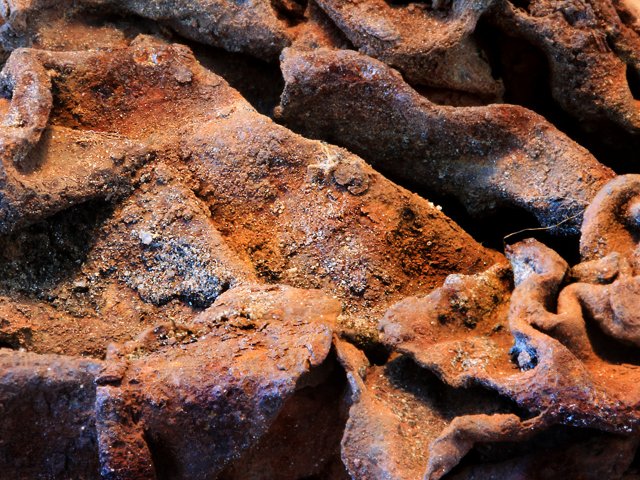}
\end{subfigure}
\begin{subfigure}{.24\linewidth}
    \centering
    \includegraphics[width=1\linewidth]{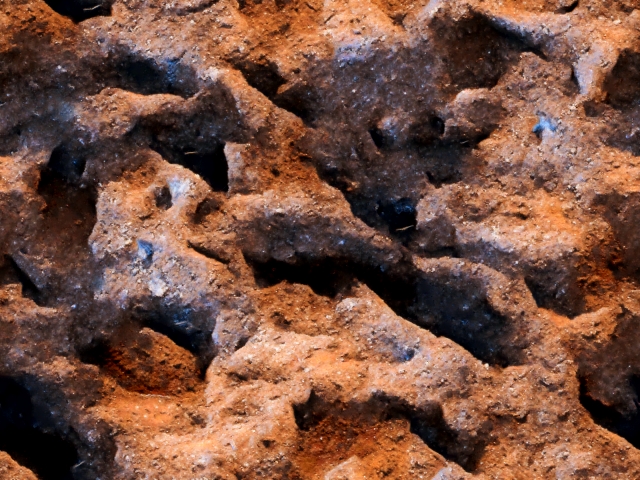}

\end{subfigure}
\begin{subfigure}{.24\linewidth}
    \centering
    \includegraphics[width=1\linewidth]{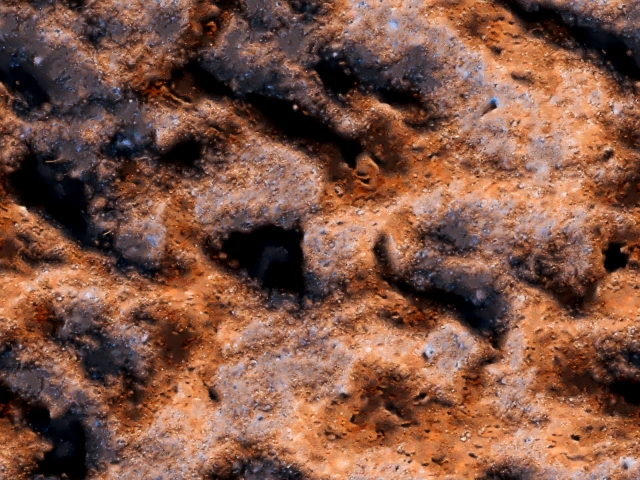}

\end{subfigure}
\begin{subfigure}{.24\linewidth}
    \centering
    \includegraphics[width=1\linewidth]{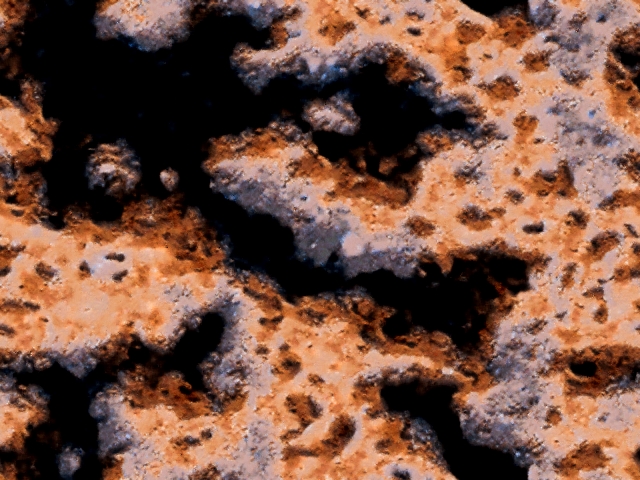}
\end{subfigure}

\caption{Entropic regularization. First image is the exemplar and subsequent images are synthesized with $\varepsilon = .001$,$.01$,$.1$ respectively. When $\varepsilon = .001$, Algorithm \ref{alg:SKP} returns a near permutation. This helps impose global image structure when synthesizing with a small patch size of 4.}\label{fig:er}
\end{figure}

In practice, small values of $\varepsilon$ reach permutations extremely quickly for most texture inputs, especially as the synthesis image approaches a permutation of the exemplar. Because matched rows and columns are removed from $\Sigma$ at each iteration, high match cardinality at each iterate greatly accelerates the algorithm.  The effect of $\varepsilon$ on image synthesis can be seen in Figure \ref{fig:er}, where lower values of $\varepsilon$ have more accurate image structure and color histograms. 

\par
Algorithm \ref{alg:LMSK} is implemented trivially on the gpu, as is the original SK algorithm, where it enjoys a significant speedup \cite{cuturi2013sinkhorn}. In fact, Algorithm \ref{alg:LMSK} and \ref{alg:SKP} are simple enough to have a fast implementation in native MATLAB, using only vectorized tensor and matrix products, calling \texttt{gpuArray} of inputs to run the algorithms on the gpu. Native MATLAB code for every algorithm and experiment in this document is provided on page 1. 
 
\section{Innovative Synthesis}
Texture synthesis is an ambiguously defined problem and ultimately depends on human observation. Even today, with the explosion of generative methods initiated by Generative Adversarial Networks \cite{goodfellow2014generative}, the success of generative methods is still largely determined by human inspection. Nevertheless, we supplement our inspection with various methods, such as the Inception Score \cite{salimans2016improved}, to provide non-visual statistical cues or confidence over large datasets for generated images. 

\begin{figure}[!ht]

\begin{subfigure}{.24\linewidth}
    \centering
    \includegraphics[width=1\linewidth]{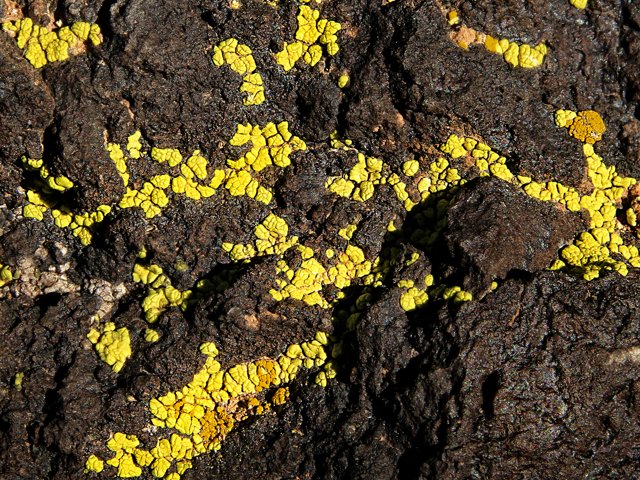}
    Exemplar
\end{subfigure}
\begin{subfigure}{.24\linewidth}
    \centering
    \includegraphics[width=1\linewidth]{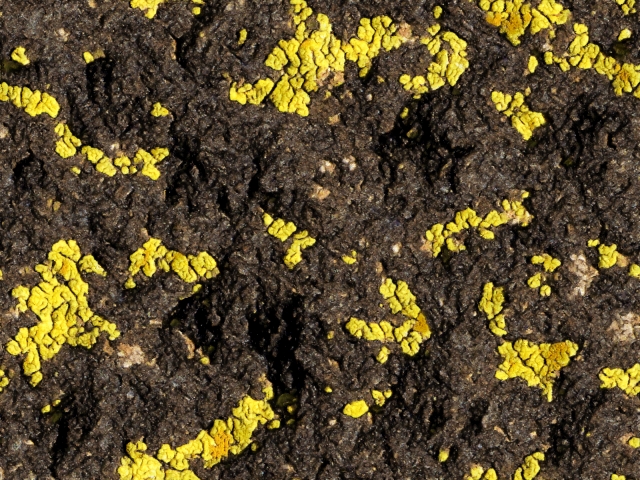}
    $IC = .81$
\end{subfigure}
\begin{subfigure}{.24\linewidth}
    \centering
    \includegraphics[width=1\linewidth]{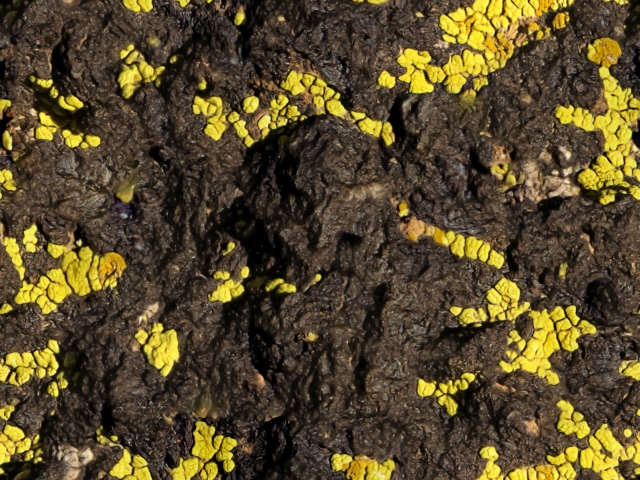}
    $IC = .43$
\end{subfigure}
\begin{subfigure}{.24\linewidth}
    \centering
    \includegraphics[width=1\linewidth]{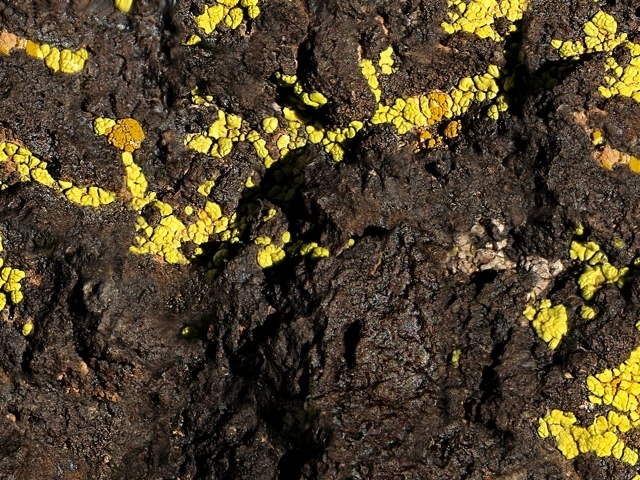}
    $IC = .33$
\end{subfigure}

\begin{subfigure}{.24\linewidth}
    \centering
    \includegraphics[width=1\linewidth]{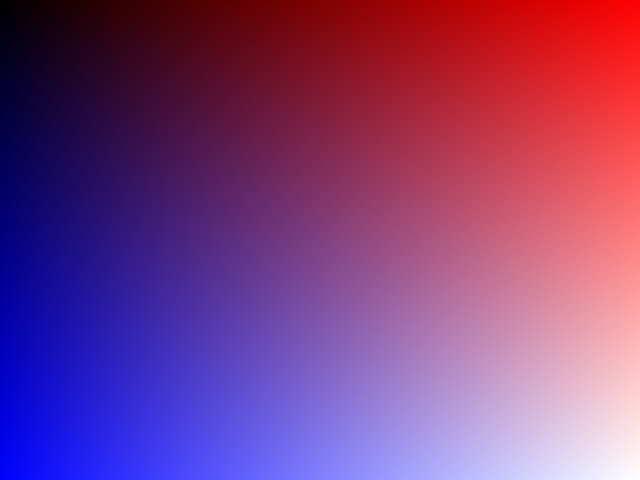}
\end{subfigure}
\begin{subfigure}{.24\linewidth}
    \centering
    \includegraphics[width=1\linewidth]{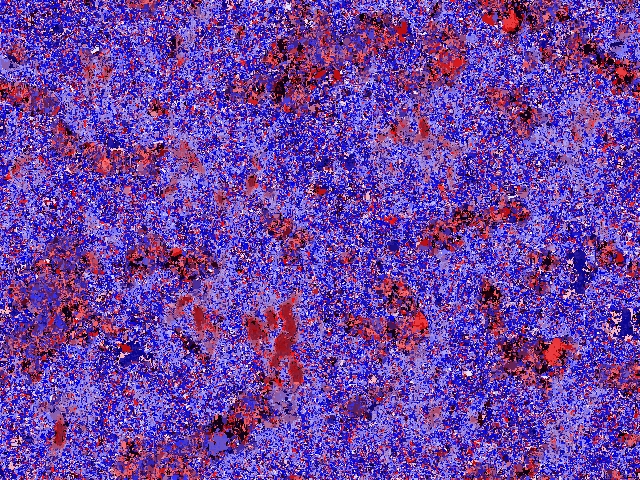}
\end{subfigure}
\begin{subfigure}{.24\linewidth}
    \centering
    \includegraphics[width=1\linewidth]{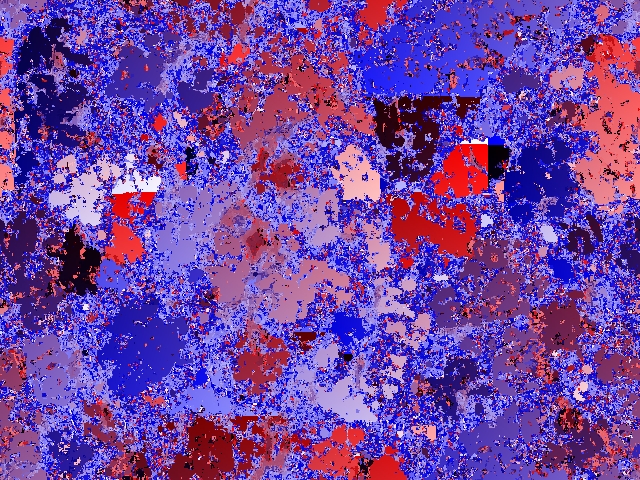}
\end{subfigure}
\begin{subfigure}{.24\linewidth}
    \centering
    \includegraphics[width=1\linewidth]{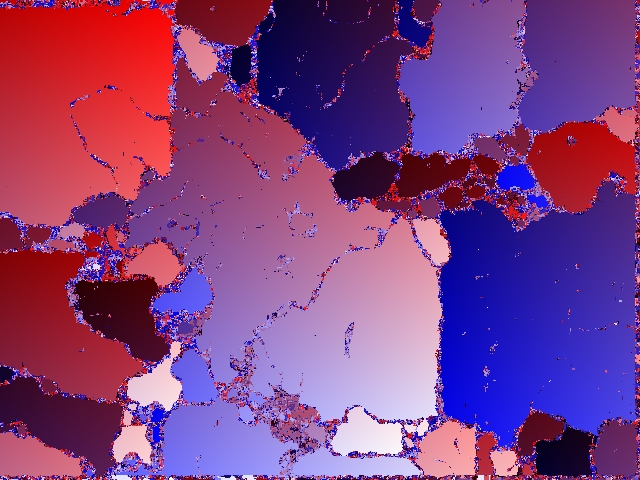}
\end{subfigure}

\caption{Innovation capacity. The second and third column were synthesized with OT $\varepsilon = .001$ and patch size 4 and 7 respectively, the fourth image with \cite{kaspar2015self} and a patch size of 4. Listed are the multi resolution innovation capacities with $J=4$ resolutions. The third image converged to a tiling at low resolution due to its large patch size. \cite{kaspar2015self} can synthesize with small patch size but its search is intrinsically limited to tilings via PatchMatch \cite{barnes2009patchmatch} . The second column, however, does not contain any salient copies which is corroborated by its high innovation capacity and tile map.}\label{fig:icps}
\end{figure}

For exemplar based texture synthesis, the ultimate goal is to be as perceptually close to the input without egregiously copying it. For example, while Self Tuning Texture Optimization \cite{kaspar2015self} can achieve a high resolution synthesis extremely quickly, its match heuristic is limited to tilings and thus it always contains salient copies of the input, see Figure \ref{fig:icps}. We propose to measure what percentage of the image was tiled as the \textit{Innovation Capacity}. Of course, euclidean distance in RGB space is sensitive to noise and diffeomorphism. To make this notion slightly more robust, we average the Innovation Capacity over every synthesis resolution in Algorithm \ref{alg:IC}.

\begin{algorithm}
\caption{Multi-Resolution Innovation Capacity}
\label{alg:IC}
\begin{algorithmic}[1]
\Input Exemplar image $x$, Synthesis $y$, Number of scales $J$
\Output Innovation Capacity $IC$
\For{$j = J \dots 0$}
\State $x_{j} \gets x$ at resolution $2^{-j}$
\State $y_{j} \gets y$ at resolution $2^{-j}$
\State $\tilde{\sigma} \gets$ Nearest neighbor match between $x_{j},y_{j}$ with lines $5\dots 8$ of Algorithm \ref{alg:MRTS}
\State $t_{x} \gets $ Identity tile map, i.e. $\{1\dots N^{2}\}$ reshaped to $N\times N$
\State $T_{x} \gets 3\times3$ patches of $t_{x}$, with center index removed. 
\State $T_{y} \gets \tilde{\sigma}T_{x}$ 
\State $IC_{j} \gets \frac{\Vert T_{x} - T_{y} \Vert_{0}}{8N^{2}}$ I.e. percentage of tiled pixels
\EndFor
\State $IC \gets$ Mean over $\{IC_{j}\}_{0\leq j \leq J}$
\end{algorithmic}
\end{algorithm}

\subsection{Comparison with Statistical Synthesis}
To give perspective of how innovative our algorithm is we compare the innovation capacity to a statistical algorithm. Modern statistical algorithms, such as \cite{gatys2015texture}, are breathtaking in their ability to synthesize textures without copying them. \cite{gatys2015texture}, however, implicitly uses millions of labeled images, as it uses the vgg-19 pre-trained network and optimizes over an enormous set of parameters. This complexity makes the algorithm somewhat uninsightful, other than gram matrices of filters are well suited for synthesis, which was known over a decade earlier \cite{portilla2000parametric}. 

As the flavor of this work is simplicity, we disregard \cite{gatys2015texture} and instead compare our algorithm to synthesis with random filters \cite{ustyuzhaninov2016texture}. This turns out to be a more appropriate comparison, as both random filter methods and our method use only the exemplar. The algorithm can be viewed as a single layer network, with a single convolution followed by a rectified linear unit. The synthesis is then optimized to minimize the distance between its gram matrix of random features and that of the exemplar. Algorithm 5 presents the objective function.

\begin{algorithm}
\caption{Gram Loss Objective}
\label{alg:RCGO}
\begin{algorithmic}[1]
\Input Current optimizer state, including current $x$ and $y$
\Output Gram matrix loss $\mathcal{L}$ and derivative $\frac{ \partial \mathcal{L}}{\partial y}$

\State $ z \gets $ Filter bank convolution of $y$ with $ b \times b \times 3 \times K $ Gaussian noise
\State $ z \gets z \otimes (z>0)$ I.e. the rectified linear unit.
\State $ z \gets $ Reshape to two dimensions, with the number of filters $K$ in dimension two
\State $G_{y} \gets z^{\intercal}z$ Gram matrix of features
\State $G_{x} \gets $ Gram matrix of $x$ computed with lines $1\dots 4$
\State $ \mathcal{L} \gets \Vert G_{y} - G_{x} \Vert _{2}^{2}$
\State $\frac{ \partial \mathcal{L}}{\partial y} \gets $ Compute with auto differentiation using back propogation
\end{algorithmic}
\end{algorithm}

We use the same multi resolution framework as \ref{alg:MRTS}, starting a new optimization at increasing resolutions. Finally, we use L-BFGS as it substantially improves the convergence times, which still takes 500 or so iterations. Algorithm 5 also has serious memory issues. This is because when too few random features are used, the optimizer quickly converges on a minimum. Synthesis is only successful with more filters. The sweet spot is around $256$ filters, which is still limits the resolution at which you can perform convolution in one shot. 

\subsection{Discussion}
\begin{figure}

\begin{subfigure}{.24\linewidth}
    \centering
    Exemplar
    \includegraphics[width=1\linewidth]{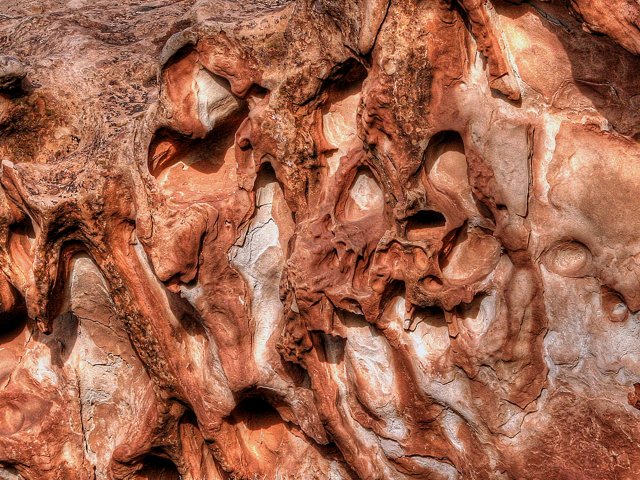}
\end{subfigure}
\begin{subfigure}{.24\linewidth}
    \centering
    OT \eqref{5} $\varepsilon = .001$
    \includegraphics[width=1\linewidth]{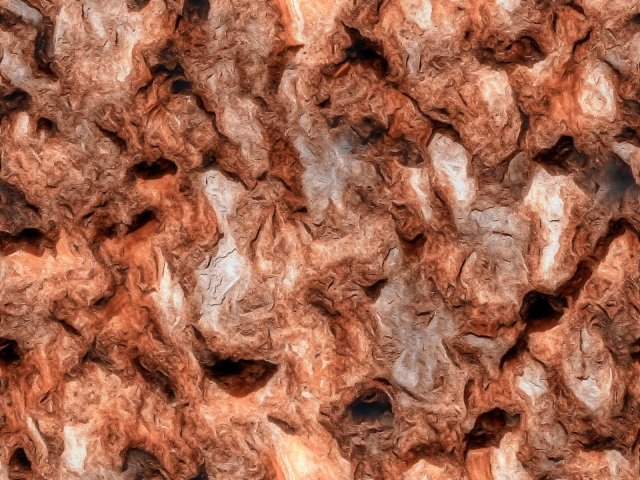}
\end{subfigure}
\begin{subfigure}{.24\linewidth}
    \centering
    BS \eqref{2} $\alpha = .25$
    \includegraphics[width=1\linewidth]{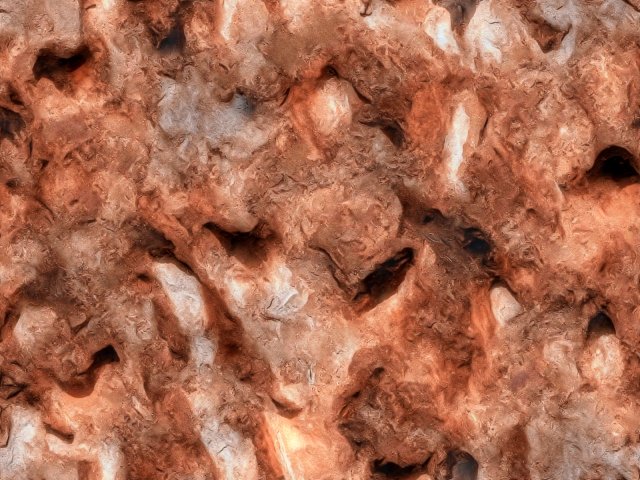}
\end{subfigure}
\begin{subfigure}{.24\linewidth}
    \centering
    Gram Loss \ref{alg:RCGO}
    \includegraphics[width=1\linewidth]{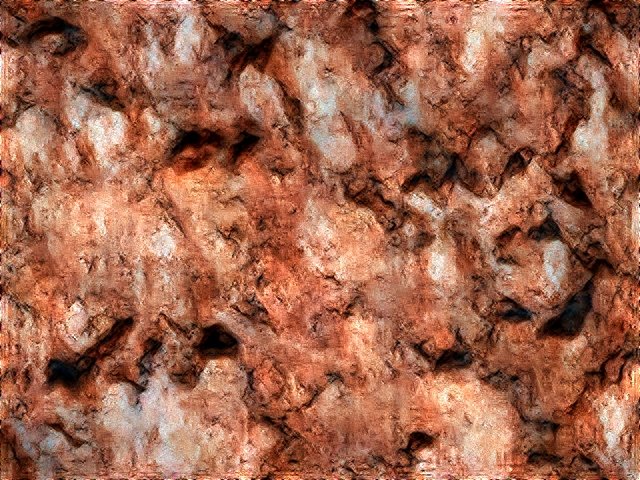}
\end{subfigure}

\medskip
\begin{subfigure}{.24\linewidth}
    \centering
    \includegraphics[width=1\linewidth]{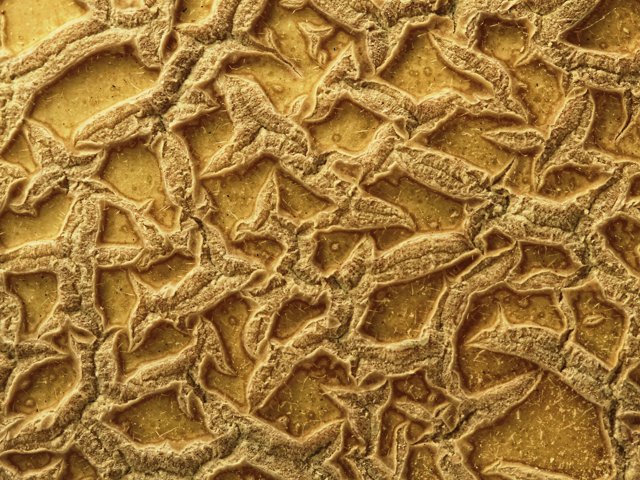}
\end{subfigure}
\begin{subfigure}{.24\linewidth}
    \centering
    \includegraphics[width=1\linewidth]{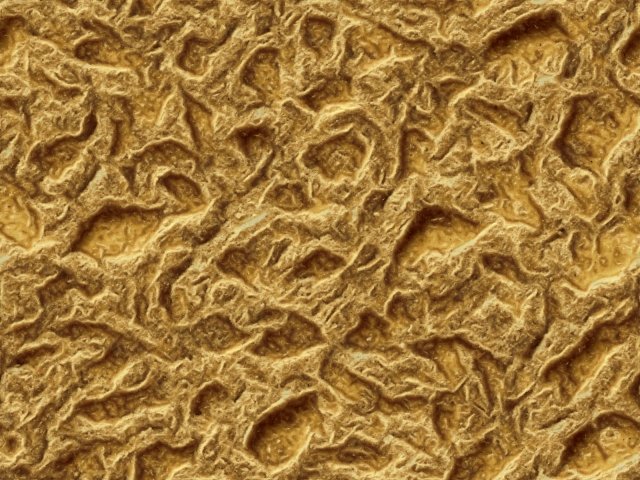}
\end{subfigure}
\begin{subfigure}{.24\linewidth}
    \centering
    \includegraphics[width=1\linewidth]{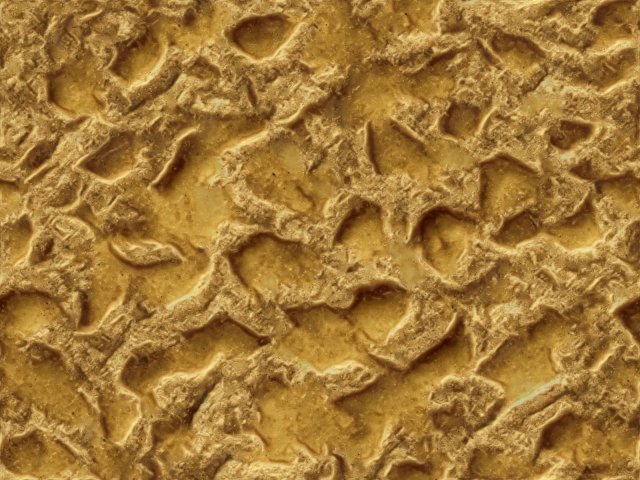}
\end{subfigure}
\begin{subfigure}{.24\linewidth}
    \centering
    \includegraphics[width=1\linewidth]{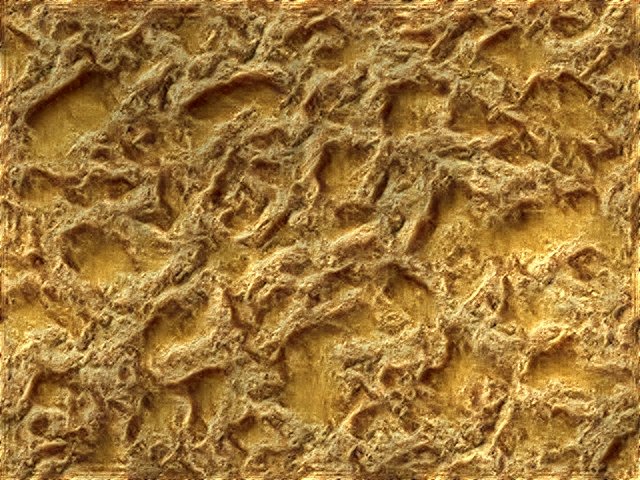}
\end{subfigure}

\medskip
\begin{subfigure}{.24\linewidth}
    \centering
    \includegraphics[width=1\linewidth]{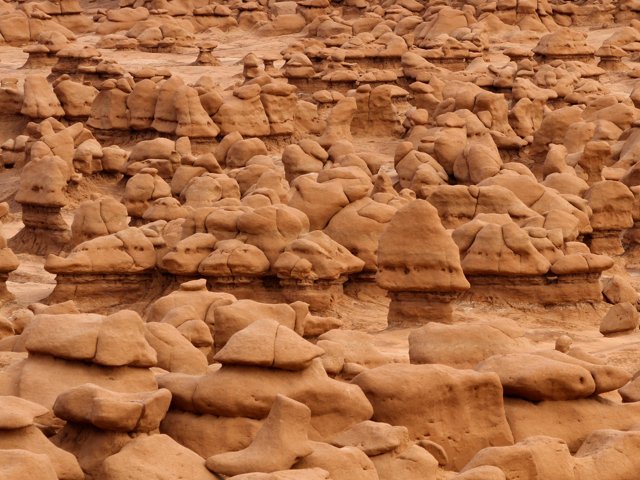}
\end{subfigure}
\begin{subfigure}{.24\linewidth}
    \centering
    \includegraphics[width=1\linewidth]{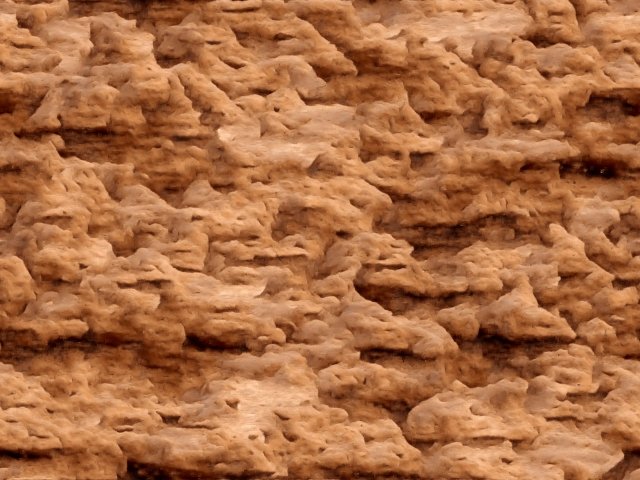}
\end{subfigure}
\begin{subfigure}{.24\linewidth}
    \centering
    \includegraphics[width=1\linewidth]{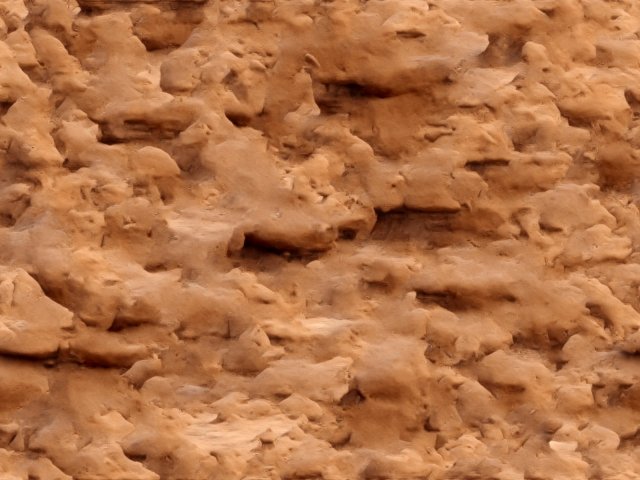}
\end{subfigure}
\begin{subfigure}{.24\linewidth}
    \centering
    \includegraphics[width=1\linewidth]{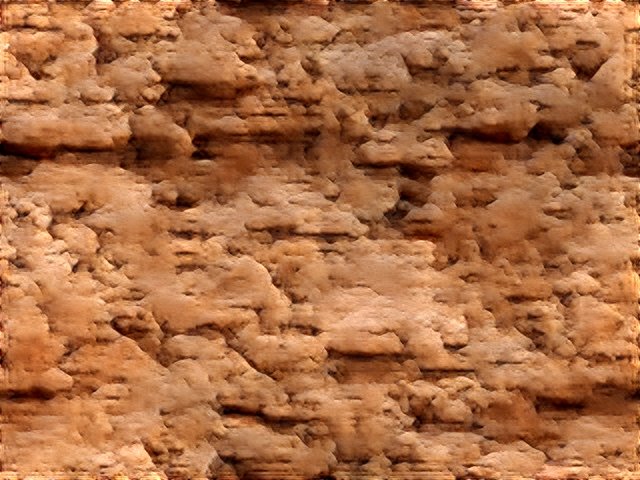}
\end{subfigure}

\medskip
\begin{subfigure}{.24\linewidth}
    \centering
    \includegraphics[width=1\linewidth]{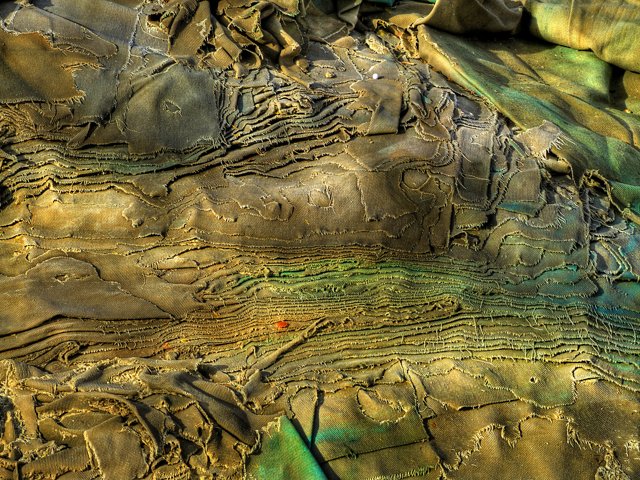}
\end{subfigure}
\begin{subfigure}{.24\linewidth}
    \centering
    \includegraphics[width=1\linewidth]{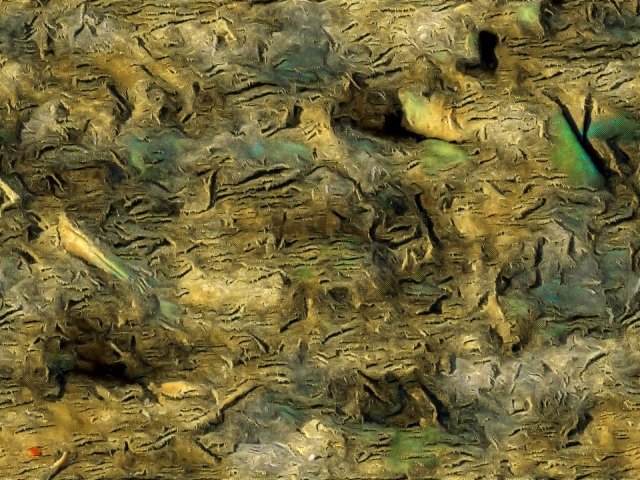}
\end{subfigure}
\begin{subfigure}{.24\linewidth}
    \centering
    \includegraphics[width=1\linewidth]{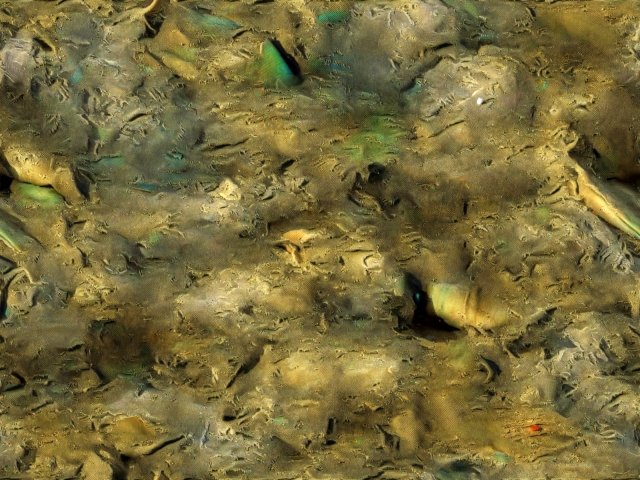}
\end{subfigure}
\begin{subfigure}{.24\linewidth}
    \centering
    \includegraphics[width=1\linewidth]{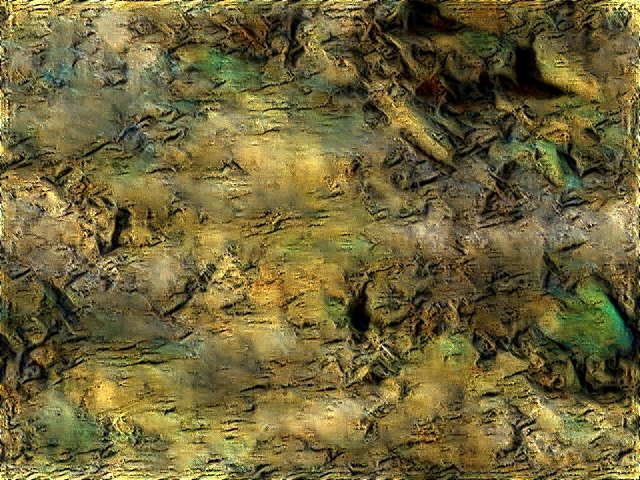}
\end{subfigure}

\medskip
\begin{subfigure}{.24\linewidth}
    \centering
    \includegraphics[width=1\linewidth]{images/lichen_lava}
\end{subfigure}
\begin{subfigure}{.24\linewidth}
    \centering
    \includegraphics[width=1\linewidth]{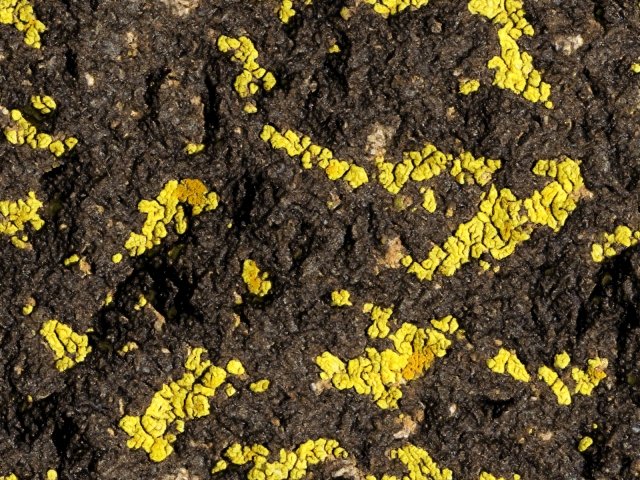}
\end{subfigure}
\begin{subfigure}{.24\linewidth}
    \centering
    \includegraphics[width=1\linewidth]{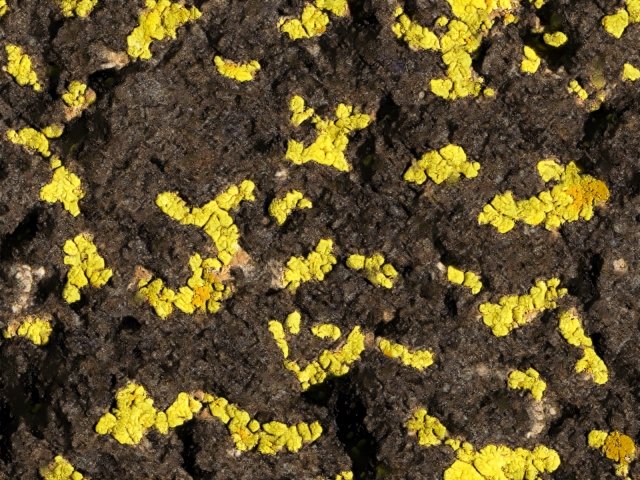}
\end{subfigure}
\begin{subfigure}{.24\linewidth}
    \centering
    \includegraphics[width=1\linewidth]{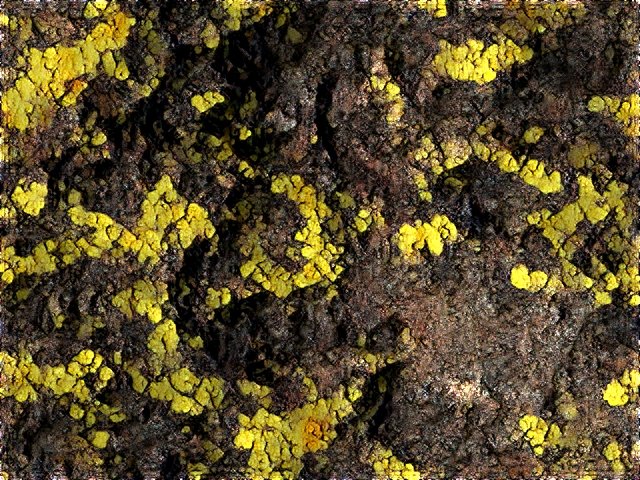}
\end{subfigure}

\medskip
\begin{subfigure}{.24\linewidth}
    \centering
    \includegraphics[width=1\linewidth]{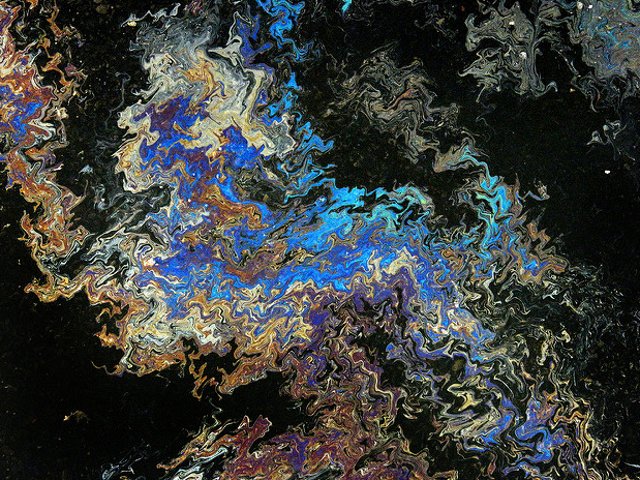}
\end{subfigure}
\begin{subfigure}{.24\linewidth}
    \centering
    \includegraphics[width=1\linewidth]{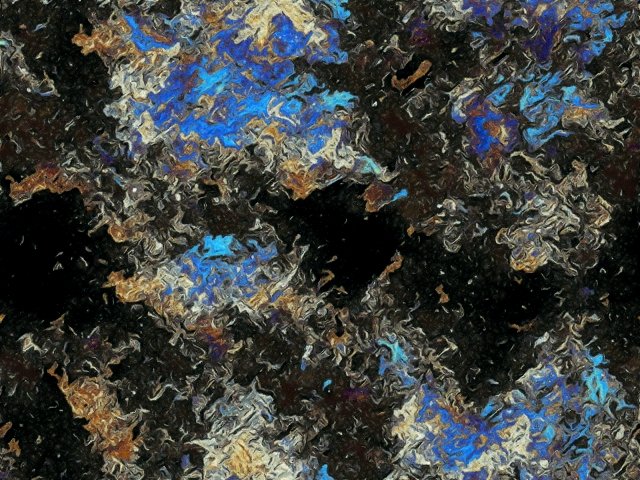}
\end{subfigure}
\begin{subfigure}{.24\linewidth}
    \centering
    \includegraphics[width=1\linewidth]{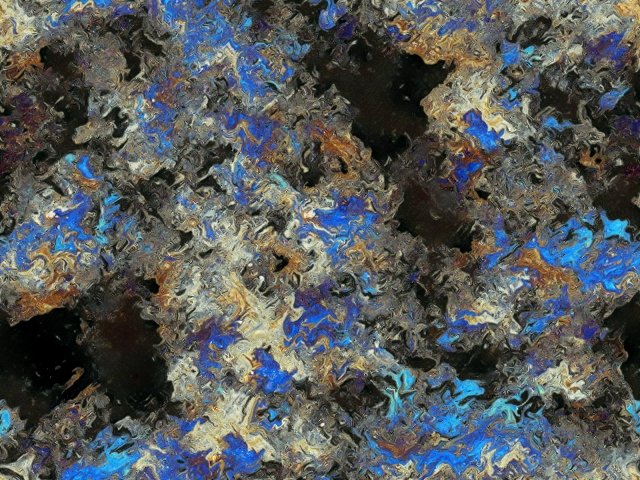}
\end{subfigure}
\begin{subfigure}{.24\linewidth}
    \centering
    \includegraphics[width=1\linewidth]{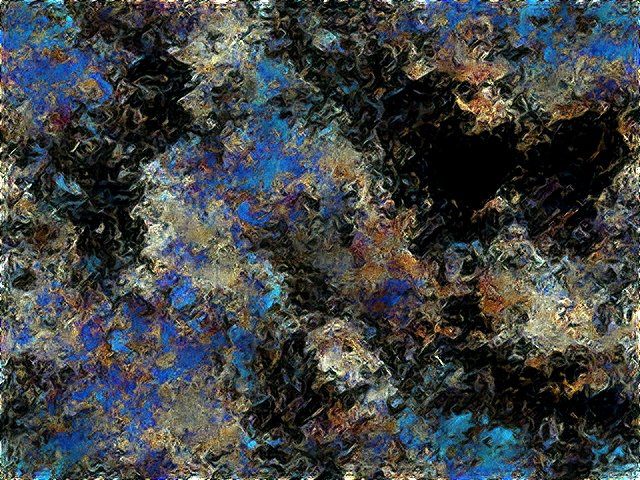}
\end{subfigure}

\medskip
\begin{subfigure}{.24\linewidth}
    \centering
    \includegraphics[width=1\linewidth]{images/red_lichen}
    Average $IC = $
\end{subfigure}
\begin{subfigure}{.24\linewidth}
    \centering
    \includegraphics[width=1\linewidth]{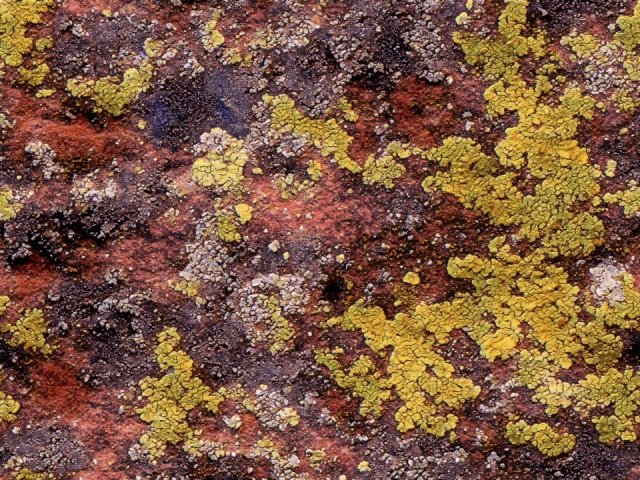}
    $.79$
\end{subfigure}
\begin{subfigure}{.24\linewidth}
    \centering
    \includegraphics[width=1\linewidth]{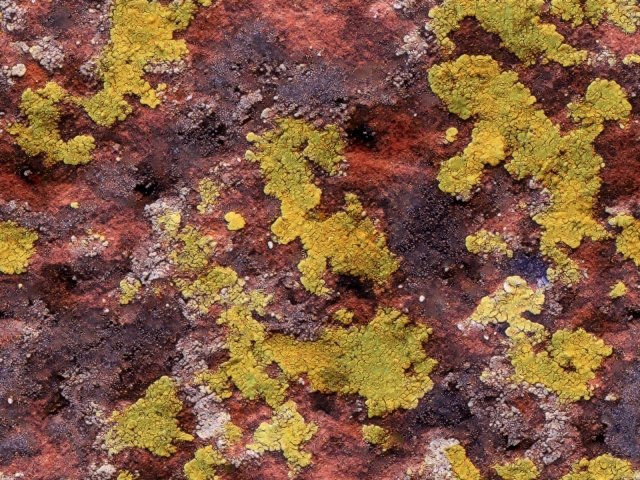}
    $.75$
\end{subfigure}
\begin{subfigure}{.24\linewidth}
    \centering
    \includegraphics[width=1\linewidth]{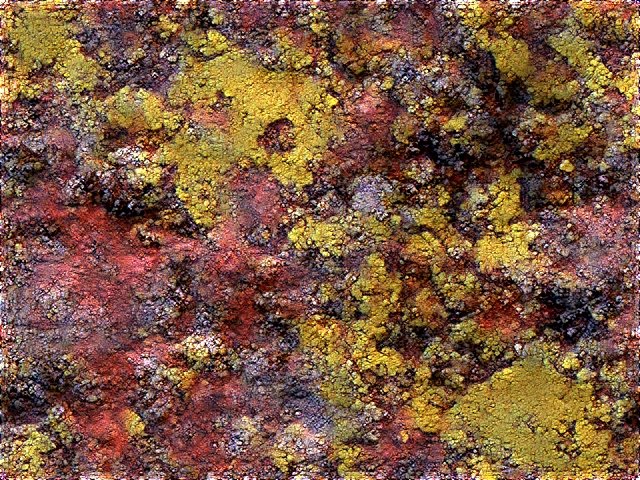}
    $.83$
\end{subfigure}

\caption{Synthesis Results. First column are the exemplars, second and third are computed with OT $\varepsilon = .001$, BS $\alpha = .25$ respectively with a patch size of 4 and fourth column is synthesis with a random filter gram loss \ref{alg:RCGO} with 256 filters. Under each column is the average multi resolution innovation capacity, computed with $J=4$ resolutions.}\label{fig:icts}
\end{figure}

Figure \ref{fig:icts} compares synthesis results of the non-parametric OT \eqref{5}, BS \eqref{2} methods and the statistical random convolution method in Algorithm \ref{alg:RCGO}. All methods are fairly plausible and contain few salient copies, which is in accordance with their high innovation capacities. We invite the reader to zoom in on the images and verify this by inspection. There are some copies present, which tend to be the non-textural pieces of the input image. For example, if the input color distribution contains a peculiarity, such as a small blotch of red, then Algorithm \ref{alg:SKP} will reach a consistent strong match. The greedy search in BS \eqref{2} was slightly more likely to verbatim copy, for example in the oil image where it obtained an innovation capacity of $IC = .67$ as opposed to $.79$ and $.83$ achieved by OT and random convolution respectively. Finally, Algorithm \ref{alg:RCGO} seems to represent the color distributions slightly better than the non-parametric methods while being noisy and representing geometric structures slightly worse. OT's advantage over BS was in the color distributions, which is not surprising as its matches are nearly permutations, while BS had a match cardinality of $MC \approx .3$ in practice. Interestingly, when the statistical method of Algorithm \ref{alg:RCGO} starts at too low of resolution or uses too many filters, the gram matrix becomes uniquely defined and Algorithm \ref{alg:RCGO} converges on a circular shift of the image. At higher resolutions, it returns a high innovation capacity because it is slightly distorted, while the egregious copies are still noticeable by inspection. This helps justify the multi resolution Innovation Capacity, as it would sometimes be extremely low for this method under certain parameters. In fact, \cite{gatys2015texture} can be optimized with far less parameters (stopping at \textit{relu3\_1}), when optimized over multiple resolutions while potentially suffering this same pitfall. Additionally, it will also be prone to noticeably copying geometric and color peculiarities in the input, even under the its original formulation. 

\subsection{Future Work}
Synthesis with the VGG-19 convolutional network \cite{gatys2015texture} is undeniably the state of the art for exemplar based texture synthesis, up to small improvements based upon that method. However, this work shows that texture synthesis algorithms that use only the input image can synthesize images fairly well and training a massive complex network on millions of labeled images is likely overkill. This is in accordance with a number of recent methods, including the Spatial GAN \cite{jetchev2016texture}, which learns a generative image representation from texture patches of the input image. Spatial GANs build the textures "from scratch", as they begin generation from a noise vector and because of this require a huge amount of parameters, typically more than the number of pixels in the image. Of course, a non-parametric algorithm needs the entire exemplar as well but images are highly compressible while convolution filters are not.
\par
The biggest pitfall to the non-parametric algorithms in this document is that euclidean distance in RGB space is unstable to diffeomorphism, especially with a larger patch size. That is, image regions should be expected to be slightly deformed to corresponding regions in the synthesis, as they are in \cite{gatys2015texture}. Convolutional networks, such as wavelet scattering networks \cite{bruna2013invariant} or the vgg-19 network, create image representations that are stable to diffeomorphism, which aids their ability to recognize the same objects under slightly different appearances. We think the simplicity of non-parametric methods is still valuable, especially when equipped with an elegant distance metric such as OT. A hybrid method may capture the best of both worlds, using a shallow learned or fixed patch representation that are stable to diffeomorphism. The challenge is fully integrating the patch representation into the texture optimization, where patch representations will need to be inverted. Perhaps one could use the recent method of SinkhornAutoDiff \cite{geneway2017learning} to provide a differentiable entropic OT loss between synthesis and exemplar representations, so that the optimization could be accomplished with SGD. This way, one has the benefit of not needing to build the image from scratch via OT and a more meaningful distance metric through the patch representation. 

\section{Conclusion}
In this work, we demonstrated that non-parametric algorithms can produce novel images. Innovative synthesis is achieved with a small patch size, where global plausibility depends on the qualities of the match. A match heuristic using entropic optimal transport was well suited for memory intensive applications such as texture synthesis. Less entropic regularization corresponded with more plausible images. Finally, we defined a metric to help determine how novel a synthesized image is. The OT and BS methods were capable of plausible synthesis with high innovation capacity, corroborated by a visual inspection revealing few or no egregious copies. 

\bibliographystyle{plainnat}
\bibliography{tsbib}

\end{document}